# SIMBOL-X: A FORMATION FLIGHT MISSION WITH AN UNPRECEDENTED IMAGING CAPABILITY IN THE 0.5-80 KEV ENERGY BAND

**Gianpiero Tagliaferri[1], Philippe Ferrando[2], Jean-Michel Le Duigou[3], Giovanni Pareschi[1], Philippe Laurent[3], Giuseppe Malaguti[4], Rodolphe Clédassou[3], Mauro Piermaria[5], Olivier La Marle[3], Fabrizio Fiore[6], Paolo Giommi[5]**

[1] INAF Osservatorio Astronomico di Brera, Via E. Bianchi 46, 23807 Merate, Italy, E-mail : gianpiero.tagliaferri@brera.inaf.it, giovanni.pareschi@brera.inaf.it
[2] CEA/DSM/IRFU/SAp & UMR 7164 Laboratoire APC, CEA-Saclay, 91191 Gif Sur Yvette Cedex, France, E-mail : philippe.ferrando@cea.fr, philippe.laurent@cea.fr
[3] Centre National d'Etudes Spatiale (CNES), 18 Avenue Edouard Belin, 31401 Toulouse Cedex 9, France, E-mail : jean-michel.leduigou@cnes.fr, rodolphe.cledassou@cnes.fr, olivier.lamarle@cnes.fr
[4] INAF IASF-Bo, Via P. Gobetti 101, 40129 Bologna, Italy, E-mail : malaguti@iasfbo.inaf.it
[5] Agenzia Spaziale Italiana (ASI), Viale Liegi 26, 00198 Roma, Italy, E-mail : mauro.piermaria@asi.it, paolo.giommi@asdc.asi.it
[6] INAF Osservatorio Astronomico di Roma, Via Frascati 33, 00040 Monteporzio, Italy, E-mail : fiore@oa-roma.inaf.it

## ABSTRACT

The discovery of X-ray emission from cosmic sources in the 1960s has opened a new powerful observing window on the Universe. In fact, the exploration of the X-ray sky during the 70s–90s has established X-ray astronomy as a fundamental field of astrophysics. Today, the emission from astrophysical sources is by large best known at energies below 10 keV. The main reason for this situation is purely technical since grazing incidence reflection has so far been limited to the soft X-ray band. Above 10 keV all the observations have been obtained with collimated detectors or coded mask instruments. To make a leap step forward in X-ray astronomy above 10 keV it is necessary to extend the principle of focusing X ray optics to higher energies, up to 80 keV and beyond. To this end, ASI and CNES are presently studying the implementation of a X–ray mission called Simbol-X.

Taking advantage of emerging technology in mirror manufacturing and spacecraft formation flying, Simbol-X will push grazing incidence imaging up to ~ 80 keV and beyond, providing a strong improvement both in sensitivity and angular resolution compared to all instruments that have operated so far above 10 keV. This technological breakthrough will open a new high-energy window in astrophysics and cosmology. Here we will address the problematic of the development for such a distributed and deformable instrument. We will focus on the main performances of the telescope, like angular resolution, sensitivity and source localization. We will also describe the specificity of the calibration aspects of the payload distributed over two satellites and therefore in a not "frozen" configuration.

## 1. INTRODUCTION

The X-ray sky was pioneered in the early '60s by R. Giacconi and his collaborators who discovered the first X-ray source, Sco X-1, beside our own Sun, and the existence of a diffuse X-ray background. With the launch of the UHURU and other satellites hundreds of X-ray sources were soon discovered, including binary stars containing compact objects, such as neutron stars and black-hole, and extragalactic active galactic nuclei (AGN). These discoveries were made with passive collimator detectors with a very poor angular resolution (of the order of tens of arcminutes). With the launch of the *Einstein* satellite we entered the realm of focussing X-ray astronomy, with a big improvement both in sensitivity and angular resolution. Since then various satellites with imaging capability have been launched. Thanks to the *Chandra* and XMM-*Newton* satellites we are now able to obtain spectacular images of the X-ray sky with the detection of hundred of thousands of sources of all kind. Unfortunately this leap step forward has been possible only below ~ 10 keV. Above this value, i.e. in the so called hard X-ray band, so far we could use only passive collimators, as those on board of the *Beppo*SAX and *Rossi*XTE satellites, or coded masks as those on board of the INTEGRAL and Swift satellites. As a result only a few hundred sources are known in the whole sky in the 10–100 keV band, a situation recalling the pre–Einstein era at lower energies. However this band is very important. For instance the X-ray background has its peak emission at 30 keV and we still do not know which sources are making up most of this emission, a question of cosmological importance. To discover them we need an instrument with much higher sensitivity and imaging capability, to separate the various weak X-ray sources. This band is also very important for the study of matter accretion around black holes and the particle acceleration mechanisms. For these reasons, there is a clear need for a high energy astrophysics mission to bridge this gap of sensitivity, i.e. to have a hard X-ray instrument with a sensitivity and angular resolution similar to those of the current soft X-ray telescopes. A hard X-ray focusing optics is needed to do this. With the emerging technology in mirror manufacturing, providing that one can achieve a very long focal length, it is now possible to develop such a mission. In particular Simbol-X, a new mission currently under study by the French and Italian agencies will have the capability to obtain X-ray images in the band 0.5–80 keV with very good sensitivity and spatial resolution [1]. Like the *Einstein* Observatory, this mission will have the capabilities to investigate almost any type of X-ray source, from Galactic and extragalactic compact objects to supernova remnant, young stellar objects and clusters of galaxies, right in the domain where accretion processes and acceleration mechanisms have their main signatures.

Simbol-X highly demanding scientific goals will be met through various technological improvements. First of all, a very long focal length, which is mandatory in order to efficiently focus the hard X-ray photons, will be obtained by using a formation flying strategy; two satellites will carry one the focusing mirrors and the other one the focal plane detector. The two satellites will be maintained at a distance of 20 m from each other, with a precision of the order of a cm [2]. Secondly, in order to efficiently reflect the X-ray photons from 0.5 to 80 keV the mirror surface will be coated with multilayers [3]. To efficiently and simultaneously detect the X-ray over such a very large energy band, two detectors, one on top of the other one, are foreseen at the focus of the mirrors [4]. Around the focal plane camera there is also an efficient anticoincidence system to obtain very low background [4]. Similarly, a pre-collimator and a proton diverter are mounted in front and behind the mirror module, respectively [5].

We will now describe more in detail, these points, including the description of the end-to-end calibrations of the two payload units (mirror module and focal plane detector) seen as a single telescope.

## 2. FORMATION FLIGHT CONCEPT AND ORBIT

The Simbol-X formation-flying concept has many aspects than can not be all described here (see [6] for a full description of the mission Global Navigation and Control (GNC) modes and the associated design). We will focus here on the observation mode and the effects of the deformable instrument concept on some of the main performances.

**2.1 General description**

During the observation mode, the mirror spacecraft (MSC) has a free flight on a high elliptical orbit with a period of about 4 days. It will point very precisely toward the source to observe. At the focal point, 20 m behind the MSC, the detector spacecraft (DSC) has a forced movement with respect to the MSC in order to maintain a suitable geometry for the science. The current status is that the center of the detector is maintained around the best mirror focus position within an accuracy of +/- 5 m m in lateral directions and +/- 3 cm in longitudinal. This was demonstrated to be feasible during the phase A using a dedicated cold gas fine propulsion module coupled with a specific optical sensor called the FFOS (Free Flying Optical Sensor). The FFOS consists in a small camera located on the DSC close to the detector payload and imaging patterns of diodes located on the mirror module. These equipments are complementary to the hydrazine propulsion and radiofrequency metrology used in the coarser acquisition and retargeting modes.

The relative position control includes a residual bias measured monthly by a dedicated calibration process involving the whole instrument pointed on a well known source. The DSC attitude control can be relaxed to +/- 2 arcmin, but the pointing knowledge has to be excellent, which implies the use a very precise star tracker (ST).

Since the controlled degrees of freedom are uncoupled a simple non-linear impulsive control is used: the dead band control allows maximizing the time between consecutive pulses while guaranteeing to stay inside the control box. The movement consists of a parabola going from the lower to the upper control bound, the relative trajectory is thus always inside a corridor whose width is tuneable depending on the mission phase. The thrust is applied against the perturbation whose magnitude is provided by the relative navigation function. The limitation on the relative control performances comes mainly from the pulse resolution. The global delta-V budget for a fine relative position control over one orbit is about 0.02 m/s and the control frequency is low (a pulse every few hundreds of seconds in the worst case).

**2.2 LOS definition and frequency classes**

One of the major topics is the restitution of the Line Of Sight (LOS) of the deformable instrument. The LOS is defined as the vector joining the center of the detector to the origin of an optical frame linked to the mirror module, expressed in an inertial frame. The in flight LOS variations are such, that the raw image formed on the detectors is blurred and has to be precisely reconstructed on ground using an estimation of the LOS associated with a very precise dating of the X-ray counts. The LOS measurement error is a complex function of sensors noises and system geometry knowledge. Full details are given in [6]. In summary, the main contributors are the DSC ST measurement errors, the FFOS errors, the relative angular position of the FFOS with respect to the DSC ST, the relative position of the FFOS components with respect to the payload elements. The LOS measurement error affects both the angular resolution of the instrument and the absolute pointing accuracy.

To separate these effects, we introduced the frequency classes reported in Tab. 1. The LOS bias is defined as the average of the LOS over one observation (few hours to 3 days). It will be calibrated each month with a residue of less than 1-2 arcsec. Due to different pointing on the vault during one month, the Sun illumination angle varies and the star trackers do not have the same pattern of stars in their field of view.

Cumulated with potential aging effects the bias has some variations between two different calibrations, mainly due to the ST field of view errors and thermoelastic effects on critical parts of the LOS. This mainly influences the absolute pointing performance.

| Constant over mission | Constant over observ. | LF over observation | HF over observation | VHF µvibrations |
|---|---|---|---|---|
| Initial bias or bias after calibration | Bias variation over 1 month ST FoVerrors, Thermo. due to sun, aging | FFOS /ST pixel noise Thermal control noise Thermal drift | ST and FFOS temporal noise Reaction wheel torque noise | Micro-vibrations |
| 0 | < $1/10/T_{obs}$ | [$1/10/T_{obs}$-$1/T_1$] | [$1/T_1$-1 Hz] | > 1 Hz |

**Table 1:** LOS frequency classes

From about $1/10^{th}$ of the observation time scale ($T_{obs}$) to the time scale of thermal or GNC control (expected to be a few hundred of seconds), the errors are of low frequency with respect to the observation. Part of it is a slow residual thermal drift after retargeting. It might appear on the reconstructed image as a deformation of the PSF in one direction. Another contribution is the thermal control induced noise.

At higher frequencies, the main contributors are the temporal noises of the sensors (read out noise, drift of stars and luminous points on the CCDs due to the satellites motion) and the residual noise of the GNC control loops. At very high frequency (typically above 1 Hz in our context), the perturbations are not measured due to the limited sampling frequency of the sensors and have a direct impact on the reconstruction process.

### 2.3 Effects of the FF on the angular resolution

We now evaluate the effect of a zero mean Gaussian LOS measurement error on the final PSF of the system, given an instrument static performance. Here we assume that the goal of a 15 arcsec HEW (see section 3) is reached at instrument level. Figure 1 plots the HEW increase as a function of the standard deviation of the LOS Gaussian noise, with various assumptions of PSF profile. It can be seen that whatever the profile, a degradation of 1 arcsec of the HEW corresponds to a 3 rms LOS measurement noise. This figure corresponds to the following set of requirements (3 sigma):

  - sensors relative alignment stability over $T_{obs}$: 1"
  - sensors temporal error at 1 Hz: 0.5"
  - FFOS spatial errors: 0.8"
  - DSC ST spatial errors: 3"
  - FFOS/payload lever of arm stab.over $T_{obs}$ : 100 µm

The LOS restitution error term has to be cumulated with the defocus term induced by the longitudinal movement of the detector around the best focus position during the observation. Current estimations show that this term is lower than a total increase of 1 arcsec of the HEW. The summation rules are not yet firmly established, but one can estimate that the total effect of the free flying on the angular resolution is less than a 2 arcsec increase with respect to the instrument static HEW.

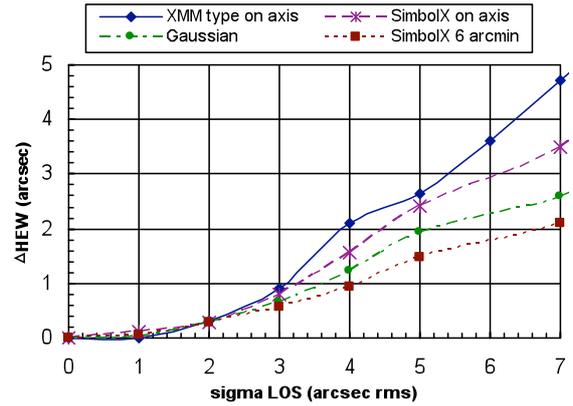

**Fig. 1 :** Increase of on axis HEW vs LOS error (15" initial HEW)

### 2.4 Absolute pointing

The absolute pointing performance is a combination of the calibration accuracy, the pixel localization in the detector payload frame and the "constant over observation" LOS measurement error class. The mission requirement leads to a set of quite demanding requirements on the LOS contributors:

- calibration of the sensor field of view error: 1.5"
- sensors relative alignment stability over 1 month: 1"
- FFOS/payload lever of arm stab. over 1 month: 50 µm

### 2.5 Other effects

The free flying concept has effects on other important performances, which are only very briefly reminded here.
First, as the instrument is open, a sky shield on the MSC and a circular baffle on the DSC are to be combined to reduce the background level, together with the anti-coincidence system around the detectors. The sizing of these two elements depends on many aspects, but in particular on the lateral and longitudinal FF tolerances. Second, the increased focal length, as compared to previous Wolter-I type instrument, leads to lower grazing incidence angles. As a consequence, the effective area of the mirror is quite sensitive to the pointing of the MSC. Some recent estimations show a variation of ~5% for a classical pointing (10 to 20 arcsec). This point is under study, but it is very likely that the mission photometry requirements will lead to a MSC pointing stability of a few arcsec and a pointing restitution at the level of accuracy of 1 arcsec.

## 3. THE MIRROR MODULE

The SIMBOL-X mirrors will be electroformed Ni shells with Wolter I profile. The adopted technology has been successfully used for the gold coated X-ray mirrors of the Beppo-SAX, XMM-Newton and Jet-X/Swift satellites. This technology has been developed and consolidated in the past two decades in Italy by the INAF Brera Astronomical Observatory in collaboration with the Media Lario Technology company. For the Simbol-X mirrors, a few important modification of the process will be implemented: 1) the use of multilayer reflecting coatings, allowing us to obtain a larger FOV and an operative range up to 80 keV and beyond; 2) the Ni walls will be a factor of two thinner than the XMM mirror shells, to maintain the weight as low as possible. With respect to the first point, once the gold-coated Ni mirror shell has been replicated from the mandrel, the multilayer film will be sputtered on the internal surface of the shell by using a two-targets linear DC magnetron sputtering system [5]. This process has been developed at the SAO–CfA for monolithic pseudo cylindrical shells [7] and now also at Media Lario where a multilayer coating facility has been developed and installed. Engineering models with two integrated shell have already been developed, with the multilayer deposition performed both at CfA and at Media Lario. In Fig.2 we show two images taken at 0.93 keV and in the band and 30-50 keV. Note how the image quality is extremely good also at the higher energies. The structure of the spider arms can be noted. The HEW at 0.93 keV is about 18 arcsec, while that one at 30 keV is ~25 arcsec, not very far away from the requirements. This somewhat higher value is due to a higher mirror surface roughness, with respect to the one required by the mission. This aspect will be investigated and improved during Phase B.

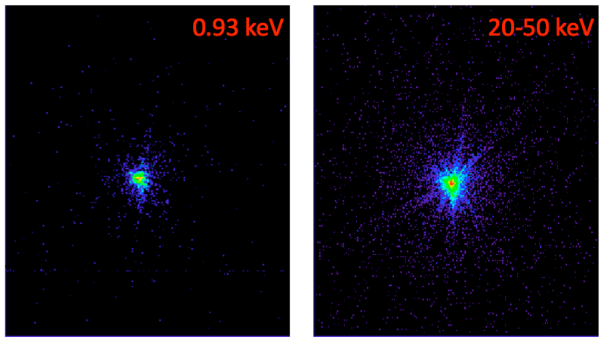

**Fig.2** X-ray images taken with a Simbol-X prototype mirror shell at the Panter facility in Munich.

### 3.1 The scientific requirements and the corresponding optics design

The Simbol-X top level requirements (see Tab. 3) translates in a number of parameters that the various part of the mission must satisfy. For the mirror module, beside the requirements on the effective area, these are:

- Angular Resolution
  $\leq$ 20 arcsec (HPD requirement) @ E < 30 keV
  $\leq$ 15 arcsec (HPD, goal) @ E < 30 keV
  $\leq$ 40 arcsec (HPD, goal) @ E = 60 keV

- Field of View
  $\geq$ 12 arcmin (diameter) @ 30 keV

These values, when compared to previous hard X-ray missions, show that Simbol-X will have unprecedented imaging capabilities. It should be noted that the very demanding requirement in terms of angular resolution is mainly dictated by the exigency of avoiding source confusion in deep surveys and to solve the puzzling problem of the hard X-ray emission from the Galactic Center region.

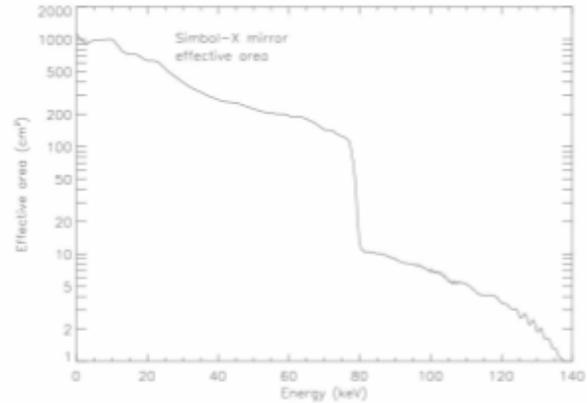

**Fig.3** Simbol-X Mirror on-axis expected effective area.

The optics configuration of Simbol-X has been addressed in the context of the Phase A study [5]. After performing a trade-off aiming at a design compliant with the requirement and compatible with the mass constraint of the mission we arrived at the set of parameters reported in Tab. 2. The mirror shells, after the integration, are kept fixed to the mechanical case by the use of two spiders, a solution already adopted for BeppoSAX and JET–X/Swift. The theoretical on-axis effective area of the mirror shell is shown in Fig. 3.

**Table 2.** Simbol-X Optics Parameters

| Number of modules | 1 |
|---|---|
| Geometrical profile | Wolter I |
| Number of nested mirror | shells 100 |
| Reflecting multilayer coating | Pt/C (200 bi-layers) |
| Focal Length | 20000 mm |
| Total Shell Height | 600 mm |
| Plate Scale | 10.3 arcsec/mm |
| Material for the mirror walls | Electroformed Ni |
| Min-Max Top Diameter | 260 to 650 mm |
| Min-Max angles of incidence | 0.18 – 0.23 deg |
| Min-Max wall thickness | 0.2 to 0.55 mm |
| Angular resolut. HEW (goal) | 20 (15) arcsec |
| Mirror shell weight | 287 kg |
| Total optics module weight | 413 kg |
| Additional mass contingency | 40 kg |

### 4. THE DETECTOR PAYLOAD

The Simbol–X detector payload is designed to achieve the scientific requirements recalled in Table 3. The parameters which are driving the design of the focal plane assembly are : i) the broad energy range, ii) the spectral resolution for the Fe and $^{44}$Ti lines energies, iii) the field of view of the optics, iv) the request for oversampling the optics point-spread function, v) the need for a very low background. This has led to use a combination of a low and high energy detector on top of each other, surrounded by a combination of active and passive shielding, as shown in Figure 4. The pixel size for each of these detectors will be $625 \times 625$ μm$^2$.

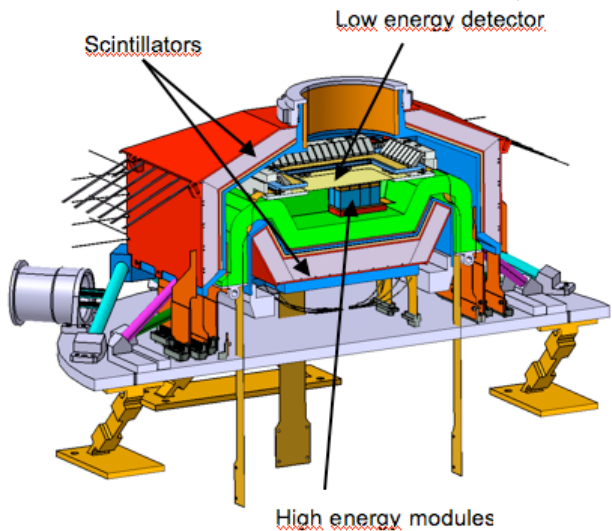

**Fig.4** : Simbol–X Focal Plane Assembly. Only part of the high energy modules are shown. The routing of the fibers connected to the scinticallator is not shown for clarity.

The low energy detector, first encountered by the beam focused by the optics, is a silicon drift detector with DEPFET (DEPleted Field Effect Transistor) readout, also called "Macro Pixel Detector" or "Active Pixel Sensor" [8]. It will cover the energy range from ~ 0.5 to ~ 20 keV. The baseline design is to cover the field of view with a single wafer, which will be logically divided into four quadrants of $64 \times 64$ pixels each. In nominal "full frame mode", all four quadrants are read out in parallel at a speed of 128 $\mu$s per frame. A faster read out is possible by reading only a part of the full area, i.e. by defining "window" modes. Timing accuracies better than ~ 50 $\mu$s can thus be achieved. Laboratory results have proven, on a prototype of a full quadrant, that the required spectral resolution is achieved at the moderately low temperatures (around –40°C) at which the instrument will be operated.

The high energy detector, less than 1 cm below the low energy one, will be constructed from 4-side juxtaposable pixelated CdTe Schottky crystals with a surface of $10 \times 10$ mm$^2$, a 2 mm thickness, and with 256 segmented electrodes. Each crystal is connected to its own read-out electronics, the IDeF-X (Imaging Detector Front-end for X-rays) ASIC. Each unit forms a complete individual X-ray camera which allows operating in the 5–100 keV range, partly overlapping the low-energy range of the silicon detector. Eight individual X–ray cameras are merged to form a 2-by-4 module having its own flex for routing signals. The detection plane will be covered with eight of such modules. As these detectors are self-triggered, sub-micro seconds timing accuracies can be achieved.

Finally, laboratory results have also shown here that the required spectral resolution can be reached at moderate temperatures, similar to those needed by the low energy detector [9].

An important requirement for sensitivity is the need to have a very low background level. In order to do this, the two imaging detectors are entirely surrounded (except for the opening to the sky) by an active anticoincidence shield coupled to a passive graded shield, which will minimize background photons produced by particles interacting in the material around and inside the detector vessel. The active shield will be made of 13 plastic scintillators tiles, of 1.5 cm thickness. These will be coupled to optical fibers, routed to multi-anode phototubes located outside of the "active volume". This is the first time that an anticoincidence system is used at soft X–ray energies, which will thus make of Simbol–X a very powerful instrument even in the soft band. This is done at the cost of a dead-time on the low energy detector, which we target to be around 30 %. The dead-time on the high energy detector is negligible. The anticoincidence scheme will be programmable.

In addition of the systems described above, forming the focal plane assembly, the detector payload also includes a calibration wheel, and a collimator. The calibration wheel has two radioactive sources (with high and low energy lines) and a closed position for protection against particle damage in belts passages and during solar flares. The collimator is a ~ 2 m long tube made of graded shield material; together with a shield located on the mirror spacecraft, it prevents the detectors to see the sky (with its diffuse X–ray emission) outside of the mirror field of view, that would give an unacceptable background level.

## 5. THE END-TO-END CALIBRATION

X-ray telescope calibration measurements are mainly devoted to the scientific characterization of the telescope, including also the production and maintenance of the calibration database (CalDB) files. Given the complex design of the Simbol-X scientific payload, its calibration will be subdivided in several phases. The first stage will foresee the subsystem functional tests of the various subsystems before integration: low energy detector, high energy detector, anticoincidence, optics shells, followed by sub-system scientific characterization. After integration, the two payloads, focal plane on one side, and mirror unit on the other one, are separately functionally tested and then scientifically calibrated. At the end of the chain, we will have the end-to-end measurement tests, with both payloads integrated and placed in a suitable calibration facility.

Instrument scientific characterization will be performed on-ground and then re-assessed in-flight. This is done

by the measurement of a number of key parameters which can be summarized as follows:

- *Effective area* ($A_{eff}$): $A_{eff}$ is the geometrical collecting area of the mirror unit convolved with its associated reflection properties and then folded with the detection efficiency of the focal plane. $A_{eff}$ depends upon various quantities the effects of which will have to be quantified during calibrations.

- *Point spread function* (*PSF*): *PSF* describes the spatial distribution of detected events in the FP in response to the observation/illumination of a point source. The canonical way to roughly quantify the *PSF* is by measuring the *Half Energy Width* (*HEW*), i.e. the diameter of the circle containing 50% of the total deposited energy.

- *Energy response*: It can be defined as the spectral energy deposit profile in response to a monochromatic incident flux. It depends primarily upon the detectors energy resolution, which is monitored by the on-board Simbol-X calibration source.

- *Timing response*: It is the capacity of locating detected events in time, with reference to an absolute clock (absolute timing accuracy), or to a detector dependent time coordinate.

- *Background*: The instrument *background* can be defined as all detected events which are not ascribed to the observed source(s). Detailed on-ground and in-flight calibration tests and observations will be devoted to the measurement of its intensity and energy spectrum, together with the assessment of its components (e.g.: stray-light, photonic diffuse, activation, soft protons, etc.).

Simbol-X novel design (separate payload in a formation flying architecture with a hybrid two-layer focal plane), and unprecedented scientific performances (three-decade wide-band deep sensitivity imaging telescope) are associated with facing novel difficulties.
*Long focal length*: the effects caused by the finite distance of the X-ray source are not anymore negligible. This means that, for a 100m long facility (the project baseline is to use the Panter test facility operated by MPE in Munich which has a X-ray beam 130m long), double reflection $A_{eff}$ loss increases up to 80-90% in the case of a full illumination set-up. This problem can have two types of solutions: **a)** the development of a dedicated pencil-beam jig tool, to be used at least for the calibration of the optics at the Panter facility; and/or **b)** the use of a longer facility (>500 m, like the one available at the Marshal-NASA facility), reaching an $A_{eff}$ loss <15%.
*Hybrid detection plane*: Simbol-X focal plane includes, for the first time at the focus of a X-ray telescope, two detector layers placed one on top of the other one,
shielded by a combination of active and passive anti-coincidence shields. Therefore, the characterization of the response of the HED will have to take into account the transparency (and scattering) caused by the LED that must be performed with very good accuracy during the laboratory tests.

## 6. CONCLUSION

The Simbol-X mission will provide an unprecedented sensitivity and angular resolution in hard X–rays, enabling to solve outstanding questions in high-energy astrophysics. To meet these goals a set of requirements, summarised in Table 3, have been defined during Phase A [10] that has demonstrated the mission feasibility.

**Table 3.** Simbol–X top-level scientific requirements

| | |
|---|---|
| Energy band | $0.5 - \geq 80$ keV |
| Field of view (at 30 keV) | $\geq 12$ arcmin (diameter) |
| On axis sensitivity | $\leq 10^{-14}$ c.g.s. 10–40 keV band, 3σ, 1 Ms |
| On axis effective area | $\geq 100$ cm$^2$ at 0.5 keV |
| | $\geq 1000$ cm$^2$ at 2 keV |
| | $\geq 600$ cm$^2$ at 8 keV |
| | $\geq 300$ cm$^2$ at 30 keV |
| | $\geq 100$ cm$^2$ at 70 keV |
| | $\geq 50$ cm$^2$ at 80 keV |
| Detectors background | $< 2\ 10^{-4}$ counts cm$^{-2}$ s$^{-1}$ keV$^{-1}$ HED |
| | $< 3\ 10^{-4}$ counts cm$^{-2}$ s$^{-1}$ keV$^{-1}$ LED |
| Line sensitivity at 68 keV | $< 3\ 10^{-7}$ ph cm$^{-2}$ s$^{-1}$, 3 σ, 1 Ms (2 10$^{-7}$ goal) |
| Angular resolution (HPD) | $\leq 20$ arcsec at E <30 keV |
| | $\leq 40$ arcsec (HPD) at 60 keV (goal) |
| Spectral resolution | E/ΔE=40–50 at 6-10 keV |
| | E/ΔE = 50 at 68 keV |
| Absolute timing accuracy | 100 μs (50 μs goal) |
| Time resolution | 50 μs |
| Absolute pointing reconstruction | $\leq 3$ arcsec (radius, 90 %) (2 arcsec goal) |
| Mission duration | 2 years of effective science time, with provision for at least 2 calendar years extension |
| Total number of pointing | 1000 (nominal mission) with provision for 500 during the 2 years extension |

We evaluated the performances of the full telescope in terms of sensitivity for the reference configuration. They are shown in Fig. 5 were we give the continuum sensitivity for measuring the broad band source spectra with 3σ accuracy for an exposure time of 1 Ms, together with the sensitivity of past and present instruments, all based on collimators or coded masks. The improvement is more than two orders of magnitude up to ~80 keV (for more details see [10]).

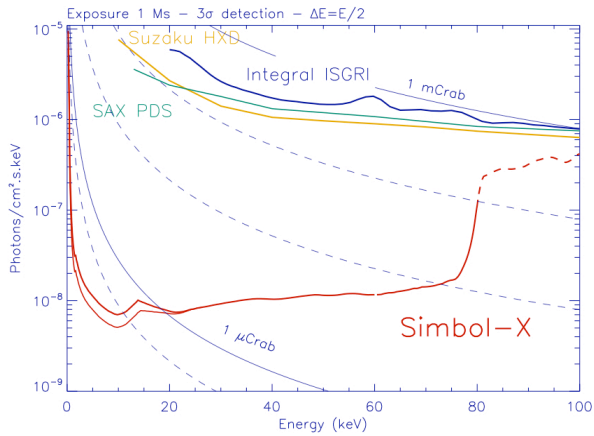

**Fig.5** Continuum sensitivity of Simbol-X for a 3σ source detection compared with previous hard X-ray telescopes.

The mission will enter in Phase B by the end of 2008 - beginning 2009, with a launch date foreseen in 2014. Simbol-X will be operated as an observatory, with a large opening to the worldwide community, and will allow breakthrough studies on black hole physics and census and particle acceleration mechanisms [11].